\documentclass[prd,reprint,superscriptaddress]{revtex4-1}

\usepackage{amsmath,amssymb}
\usepackage{verbatim}
\usepackage{graphicx}
\usepackage{hyperref}
\usepackage{color}
\usepackage{mathrsfs}
\usepackage{slashed}

\newcommand {\cK}{{\cal K}}
\newcommand {\cL}{{\cal L}}

\newcommand {\cN}{{\cal N}}

\def\a{\alpha}

\def\d{\delta}
\def\e{\epsilon}
\def\f{\phi}
\def\g{\gamma}

\def\l{\lambda}
\def\m{\mu}
\def\n{\nu}

\def\p{\pi}

\def\s{\sigma}

\newcommand{\vf}{\varphi}

\newcommand{\be}{\begin{equation}}
	\newcommand{\ee}{\end{equation}}
\newcommand{\bea}{\begin{eqnarray}}
	\newcommand{\eea}{\end{eqnarray}}

\newcommand{\ba}{\begin{array}}
	\newcommand{\ea}{\end{array}}

\def\double #1{#1{\hbox{\kern-2pt $#1$}}}

\newcommand{\bsubeq}{\begin{subequations}}
	\newcommand{\esubeq}{\end{subequations}}

\def\rmi{{\rm i}}

\def\a{\alpha}

\def\g{\gamma}

\def\d{\delta}

\def\e{\epsilon}

\def\f{\phi}
\def\vf{\varphi}

\def\p{\psi}

\def\l{\lambda}

\def\m{\mu}
\def\n{\nu}

\def\s{\sigma}

\begin{document}
	
	\title{ Carrollian Supersymmetry and SYK-like models}
	
	\author{Oguzhan Kasikci}
	\email{kasikcio@itu.edu.tr}
	\affiliation{Department of Physics,
		Istanbul Technical University,
		Maslak 34469 Istanbul,
		T\"urkiye}
	
	\author{Mehmet Ozkan}
	\email{ozkanmehm@itu.edu.tr}
	\affiliation{Department of Physics,
		Istanbul Technical University,
		Maslak 34469 Istanbul,
		T\"urkiye}
	
	\author{Yi Pang}
	\email{pangyi1@tju.edu.cn}
	\affiliation{Center for Joint Quantum Studies and Department of Physics,\\
		School of Science, Tianjin University, Tianjin 300350, China \\}
	
	\author{Utku Zorba}
	\email{zorba@itu.edu.tr}
	\affiliation{Department of Physics,
		Istanbul Technical University,
		Maslak 34469 Istanbul,
		T\"urkiye}

	\date{\today}
	
	%	\preprint{}

	\begin{abstract}
		
		This work challenges the conventional notion that in spacetime dimension higher than one, a supersymmetric Lagrangian invariably consists of  purely bosonic terms, purely fermionic terms, as well as boson-fermion mixing terms. By recasting a relativistic Lagrangian in terms of its non-relativistic and ultra-relativistic sectors, we reveal that an ultra-relativistic (Carrollian) supersymmetric Lagrangian exhibits an exotic feature, that is, it can exist without a purely bosonic contribution. Based on this result, we demonstrate a link between higher-dimensional Carrollian and (0+1)-dimensional quantum mechanical models, yielding higher-order extensions of supersymmetric SYK models in which  purely bosonic higher order terms are absent. Given that supersymmetry plays an essential role in improving the quantum behavior and solubility, our findings may lead to interesting applications in non-AdS holography. 
		
	\end{abstract}
	
	%	\pacs{??? ... ???}
	
	\maketitle
	\allowdisplaybreaks
	
	% \textit{Introduction} --  
	Supersymmetry, as a symmetry under the exchange of bosons and fermions, implies that typically a supersymmetric theory decomposes into three parts: a purely bosonic part, a purely fermionic part and a set of mixing terms that couples the fermionic and the bosonic sectors to each other. In many known examples, this statement appears to hold generically in diverse dimensions higher than one with any number of supercharges and any number of derivatives. In this work, we question this statement by decomposing a relativistic Lagrangian into its Galilean (non-relativistic) and Carrollian (ultra-relativistic) sectors. Our analysis underscores the pivotal role played by the non-relativistic sector organized according to the powers of the speed of light. Most notably, we show that it is possible for a supersymmetric ultra-relativistic Lagrangian to exist without a strict requirement for a purely bosonic part.
	
	The ultra-relativistic corner of the Bronstein hypercube has been under intense investigation in recent years, expanding our knowledge beyond the standard framework of relativistic physics. This endeavor has revitalized Carrollian physics \cite{LLJM1,LLJM2,Henneaux:1979vn} in many areas, including dynamics of null hypersurfaces \cite{Duval:2014uva,Donnay:2019jiz,Ciambelli:2019lap,Grumiller:2019fmp,Penna:2018gfx,Freidel:2022bai,Redondo-Yuste:2022czg,Marsot:2022qkx}, celestial holography and conformal Carroll holography \cite{Hartong:2015usd,Bagchi:2016bcd,Donnay:2022aba,Bagchi:2022emh,Donnay:2022wvx,Donnay:2022wvx,Bagchi:2023fbj,Saha:2023hsl}, condensed matter systems \cite{Bidussi:2021nmp,Marsot:2022imf,Bagchi:2022eui,Huang:2023zhp,Figueroa-OFarrill:2023qty,Zhang:2023jbi}, hydrodynamics \cite{deBoer:2017ing,Ciambelli:2018xat,Ciambelli:2018wre,deBoer:2023fnj,Bagchi:2023ysc,Armas:2023dcz,Bagchi:2023rwd}, Carrollian gravity \cite{Bergshoeff:2017btm,Hansen:2021fxi,Grumiller:2020elf}, black holes \cite{Ecker:2023uwm}, supergravity \cite{Ravera:2019ize,Ali:2019jjp,Ravera:2022buz} and string theory \cite{Bagchi:2016yyf,Bagchi:2017cte,Bagchi:2018wsn}. Based on the successful experience of relativistic supersymmetry, Carroll supersymmetry is expected to be a vital component of various studies for a number of reasons. It improves the quantum behavior of non-supersymmetric models and offers more tools for solving a model. Nevertheless, it is poorly understood at this stage as most studies focus on two-derivative scalar field theories and their quantization \cite{Chen:2023pqf, Banerjee:2023jpi, Figueroa-OFarrill:2023qty, deBoer:2023fnj,Koutrolikos:2023evq}.
	
	Another perspective to better understand the Carroll supersymmetry comes from the fact that a higher dimensional Carrollian model without spatial derivatives can be viewed as a collection of infinitely many quantum mechanical models, each labelled by its spatial coordinates. Consequently, it appears that Carroll symmetry can also emerge in quantum mechanical systems. From this viewpoint, the $(0+1)$-dimensional Sachdev-Ye-Kitaev (SYK) \cite{1993PhRvL..70.3339S, Kitaev} or SYK-like models \cite{Fu:2016vas,Anninos:2016szt,Murugan:2017eto,Biggs:2023mfn} are naturally related to Carrollian models \footnote{This is different from higher dimensional generalization of SYK model \cite{Cai:2017vyk} where interactions between nearest sides are turned on.}. As we will show, the lack of a purely bosonic part simultaneously leads to the absence of spatial derivatives in supersymmetric Carrollian models. Thus our result unveils another connection between Carrollian models and quantum mechanical models. In particular this means that one can obtain extensions of the recently studied supersymmetric SYK-like models as a by-product of higher-dimensional models. 
	
	% since the Carrollian supersymmetric example does not contain spatial derivatives
	
	% \textit{A Critical Look at Relativistic Supersymmetry} --  
	% We start our discussion with some general remarks and statements on relativistic supersymmetry and its Galilean limit. 
	We start our discussion by fixing our relativistic notation as $x^\mu = \left(ct, x^i\right)$ along with $\eta = \mathrm{diag}(-\,,+\,,\dots\,,+)$ where $\eta$ is the Minkowski metric. Assuming that in a relativistic Lagrangian the speed of light $c$ only appears via time derivatives (there are no other $c$-dependent parameters), and consequently when writing out the $c$ dependence explicitly, the Lagrangian is of the form $\cL = \sum_{n=0}^N c^{-n} \cK_{n}$, where $\cK_{n}$ represents the Lagrangian that is $n$-th order in time derivatives and $N$ denotes the highest number of time derivatives. Thus, a purely spatial derivative part of a Lagrangian comes with no speed of light factor, hence surviving in the $c \to \infty$ limit, which is the so-called Galilean limit of a relativistic theory. 
	
	Recall that the structure of the supersymmetry generator is of the following form,
	\begin{eqnarray}
		Q_\a = \frac{\partial}{\partial\bar\theta^\alpha} + \frac1{4 c} \left(\gamma_0 \theta \right)_\a \partial_t - \frac14 \left(\gamma^i \theta \right)_\a  \partial_i \ ,
		\label{Qtransform}
	\end{eqnarray}
	which can be recollected according to powers of $1/c$ as $\d_Q = \delta_Q^G + \d_{Q}^\prime/c$ where the superscript $G$ refers to the Galillean part that survives in the $c \to \infty$ limit.  The invariance under relativistic supersymmetry (up to total derivative) then imply 
	%%%%%%
	\be
	0=\d_Q \cL = \frac{\d_Q^\prime \cK_{N}}{c^{N+1}}  + \sum_{n=0}^{N-1}  \frac{ \delta_Q^\prime \cK_{n}  +  \d_{Q}^G \cK_{n+1}}{c^{n+1}}  + \d_Q^G \cK_0 \,.
	\label{LagCExp}
	\ee
	%%%%%%%%%%%
	This structure indicates that $\cK_N$ is invariant under $ \d_{Q}^\prime$ while $\cK_0$ is invariant under $\delta_Q^G$, and transformations of other $\cK_n$s ought to cancel with each other. As a  concrete example, consider the relativistic Lagrangian for a 3D $\cN = 1$ scalar multiplet with the field content $(\phi,\,\psi,\,f)$
	\begin{equation}
		\cL_{S} = \frac{1}{2c^2} \dot\f^2 + \frac1{8c} \bar\p \g_0 \dot\p  
		- \frac12 \partial_i \f \partial^i \f - \frac18  \bar\p  \g^i \partial_i \p  + \frac18 f^2 \,,  
		\label{2DerivativeAction}
	\end{equation}
	which is invariant under the supersymmetry transformation rules
	\begin{align}
		\d f &= \frac1{2c}\bar\e \g_0 \dot\p - \frac12 \bar\e \g^i \partial_i \p \,, & \d \f &= \frac14 \bar\e \p \,,\nonumber\\
		\d \p &= - \frac{1}{c} \g_0 \dot{\f} \e + \g^i \partial_i \f \e - \frac12 f \e
		\,.
		\label{2derivativeScalar}
	\end{align}
	Here,  $\f$ and $f$ are real scalar fields while $\psi$ is a two-component Majorana spinor. Obviously, the $\mathcal{O}(c^{-2})$ Lagrangian is invariant under $\d_Q^\prime$ transformation rules since $\d_Q^\prime \f = 0$. Similarly, the $\mathcal{O}(c^{0})$ Lagrangian, which is the Galilean supersymmetric model, is invariant under $\d_Q^G$ transformations. Finally, note that in the $c \to \infty$ limit, the time derivatives drop out from the transformation rules which is expected since the supersymmetry algebra is of the form
	\begin{eqnarray}
		\{Q,Q\} \sim - \frac{1}{c} \g_0 H +  \gamma^i P_i \,,
		\label{QQ}
	\end{eqnarray}
	indicating that in the Galilean limit, supercharges square to the spatial translations. Before finishing our discussion on Galilean models, we point out another important property of the expansion \eqref{LagCExp}. Writing out the ${\cal O}(c^{-1})$ term in the expansion of the Lagrangian, we obtain
	\begin{eqnarray}
		\d_Q \cL &=& \frac{\d_Q^\prime \cK_{N} }{c^{N+1}} + \sum_{n=1}^{N-1} \frac{ \delta_Q^\prime \cK_{n} +  \d_{Q}^G \cK_{n+1} }{c^{n+1}}   \nonumber\\
		&& + \frac{1}{c}  \left( \delta_Q^\prime \cK_{0} +  \d_{Q}^G \cK_{1} \right)  + \d_Q^G \cK_0 \,.
		\label{LagCExp2}
	\end{eqnarray}
	This form of the expansion indicates that  under supersymmetry transformation, the invariance of the action is satisfied in each order in $1/c$-expansion, independently. Consequently, if $\cK_0$ vanishes identically, then $\cK_1$ becomes Galilean invariant. 
	%Nevertheless, as $\cL_0$ contains spatial derivatives, it doesn't seem possible to obtain identically vanishing models. However, 
	If that happens, we multiply the entire Lagrangian with a factor of $c$ before taking $c \to \infty$ limit to obtain the Galilean invariant model described by $\cK_1$. As suggested by the example \eqref{2DerivativeAction}, the  action $\cK_1$ contains only fermions but no bosonic partners. Thus, the vanishing of $\cK_0$ appears to indicate that the Galilean supersymmetric model can exist without referring to any purely bosonic terms. Nevertheless, as $\cK_0$ contains only spatial derivatives but no time derivatives, it seems unlikely to obtain an identically vanishing $\cK_0$. 
	The reason is that a Lorentz invariant Lagrangian, if being non-zero, always has a non-relativistic limit with purely spatial derivatives, hence the vanishing of $\cK_0$ implies the vanishing of the original relativistic model itself \footnote{Exceptional cases may arise when the indices of the Levi-Civita tensor are contracted with derivatives, in which case one ends up with a single term in the Lagrangian with an explicit $c$-factor. Such models are relativistic by themselves, and we exclude those in our analysis.}.
	
	To be concrete, consider the following example \cite{Baig:2023yaz}
	\begin{eqnarray}
		\cL \sim \phi (\Box \phi \Box \phi - \partial_\mu \partial_\nu \phi \partial^\mu \partial^\nu \phi) \,.
		\label{example}
	\end{eqnarray}
	The $\cK_0$ part of the action is non-vanishing as the $c \to \infty$ limit only replaces the Lorentz indices with spatial indices, i.e., $\cK_0 \sim \phi (\partial^i \partial_i \phi \partial^j \partial_j \phi - \partial_i \partial_j \phi \partial^i \partial^j \phi)$. Note, however, that model has an identically vanishing $\cK_4$ as terms with highest time derivatives cancel with each other. Thus, we conclude that while $\cK_n$ for $n > 0$ might be identically vanishing, the Galilei model given by $\cK_0$ is usually nonvanishing. 
	
	The vanishing of the $\cK_4$ in our example is an encouraging signal to work out the structure of the opposite end, which is the Carrollian supersymmetry arising in the $c \to 0$ limit of relativistic supersymmetry. The algebra \eqref{QQ} suggests that we need to rescale the supersymmetry generators by a factor of $1/\sqrt{c}$, i.e., $Q \to   1/\sqrt{c}\, Q$ \cite{Bergshoeff:2015wma}. Furthermore, we need to rescale the fields with certain powers of $c$ such that neither the transformation rules, nor the Lagrangian diverges at $c \to 0$ limit. For instance, for the 3D $\cN = 1$ scalar multiplet model \eqref{2DerivativeAction}, the $c$-scaling can be assigned as follows
	\begin{align}
		\f & \to c \f\,, & \p &\to \sqrt{c} \p \,, & f & \to f\,, & \e & \to \sqrt{c} \e \,,
		\label{Scaling}
	\end{align}
	so that supersymmetry generator squares to the time translation in the  $c \to 0$ limit. The resulting Lagrangian contains only time derivatives which is invariant under the Carrollian boosts, i.e., $ x^i \to  x^i \,, t\to t+ b_i x^i$. More generically, after rescaling various fields and transformation rules, we have
	%clearly Carroll invariant.
	\begin{eqnarray}
		\cL &=& \sum_{n=0}^N c^n \cL_{n} \,, \qquad \d_Q = \d_Q^C + c \d_Q^\prime  \,.
		\label{carexp}
	\end{eqnarray}
	Note that the terms with $N$-th order time derivative have been rescaled properly so that the leading term in $\cL$ comes out as $\cL_0$, and the $c \to 0$ limit yield a finite result. The $\cL_0$ consists of terms that, in the original Lagrangian were $\mathcal{O}(c^{-N})$ and possible $\mathcal{O}(c^{-n})$ ($n<N$) terms which acquire the same powers of $c$ after rescaling the fields. Therefore, $\cL_0$ would contain the time derivatives of the bosonic fields as well as derivatives of fermions, and terms involving auxiliary fields while its purely spatial derivative part appears at  ${\cal O}(c^{N})$. To see the structure of the supersymmetry generator in the Carroll limit, we 
	first rescale $\theta \to \sqrt{c} \theta$ in \eqref{Qtransform}. Combining with rescaling of parameter $\e$ as given in \eqref{Scaling}, one finds that the supersymmetry transformation acting on a superfield $\Phi$ has precisely the structure in \eqref{carexp}
	\begin{eqnarray}
		\d_Q \Phi=  \left[\bar\e Q, \Phi \right] = \left[\bar\e Q^C, \Phi \right]  + c \left[\bar\e Q^\prime, \Phi \right]   \ .
	\end{eqnarray}
	Here, we assume that the structure of a superfield is not deformed due to rescaling the fields, and the entire superfield scales with some power of $c$. In the case of scalar  multiplet, the rescaling of the superfield $\Phi= \f + {\rm i}\bar\theta \psi + \frac{\bar\theta\theta}{2\rm i} f$, is given by $\Phi \to c \Phi$ according to \eqref{Scaling} and $\theta \to \sqrt{c} \theta$. The invariance of the Lagrangian now implies that $\delta \mathcal{L} = 0$, i.e.
	\begin{equation} 
		0 = \d_Q^C \cL_0 + \sum_{n=1}^{N}  c^{n} \left(\d_Q^C \cL_{n} + \d_Q^\prime \cL_{n-1} \right) +  c^{N+1} \d_Q^\prime \cL_N \,.
	\end{equation}
	As in the case of \eqref{example}, the contribution to $\cL_0$ from $\mathcal{O}(c^{-N})$ may be identically vanishing. In this case, the resulting Carroll invariant models does not necessarily contain a purely bosonic term. Instead, it could consists of terms that are either purely fermionic or fermions coupled to derivative of bosons. 
	%As we see momentarily that  a Carroll supersymmetric Lagrangian  without a purely bosonic part do exist. 
	
	%Unlike the Galilei case, the leading term $\cL_0$ in the small $c$ expansion the contribution may be identically 0. When this happens, $\cL_1$ gives rise to a Carroll invariant model which does not necessarily contain purely bosonic term. Instead it could consists of terms that are either purely fermionic or fermions coupled to derivative of bosons. As we see momentarily that  a Carroll supersymmetric Lagrangian  without a purely bosonic part do exist. 
	
	%\textit{An Example} -- 
	As an illustrative example with a supersymmetric Carrollian action that does not contain a purely bosonic part, we consider the recently proposed spacetime subsystem symmetric model \cite{Baig:2023yaz}, which has recently been rediscovered as Carroll swiftons that allows propagation at a non-vanishing velocity \cite{Ecker:2024czx}
	\begin{eqnarray}
		\cL_{\rm sub} &=&  \frac12 \dot \f_1^2 + \frac12 \dot \f_2^2 + \frac{\a}2 (\dot \f_1 \partial_i \f_2 - \dot \f_2 \partial_i \f_1 )^2 \,,
	\end{eqnarray}
	whose relativistic origin is given by \cite{Kasikci:2023tvs}
	\begin{eqnarray}
		\cL &=& - \frac12 \partial_\m \f_1 \partial^\m \f_1  - \frac12 \partial_\m \f_2 \partial^\m \f_2  - \frac{\a}{4} F_{\m\n} F^{\m\n}
		\,,
	\end{eqnarray}
	where $F_{\m\n} = \partial_\m \f_1 \partial_\n \f_2 -  \partial_\n \f_1 \partial_\m \f_2$.
	The two-derivative part has a well-defined limit which can be read off from \eqref{2DerivativeAction} after rescaling according to \eqref{Scaling} followed by the $c \to 0$ limit. The four-derivative terms can be supersymmetrized by considering the vector multiplet action
	\begin{eqnarray}
		\cL_{V}= - \frac{1}4 F_{\m\n} F^{\m\n} - 2  \bar\vf \slashed{\partial} \vf  \,.
		\label{VectorAction}
	\end{eqnarray}
	which is invariant under the transformation rules $\d A_\m = - \bar\e \g_\m \vf $ and $\d \vf = \frac18\, \g^{\m\n} F_{\m\n} \e ,$ where $F_{\m\n} = 2\partial_{[\m} A_{\n]}$. Based on the transformation rules of the scalar multiplet \eqref{2derivativeScalar} and the vector multiplet, the fields of the vector multiplet can be realized as composites built from the scalar multiplet
	\begin{eqnarray}
		\vf &=& \frac18 (\slashed{\partial}\f_1  \p_2 - \slashed{\partial}\f_2  \p_1 - \frac12 f_1 \p_2 + \frac12 f_2 \p_1 ) \,,\nonumber\\
		F_{\m\n} &=& \partial_\m \f_1 \partial_\n \f_2 - \partial_\n \f_1 \partial_\m \f_2  + \frac18 \bar\p_1 \g_\m \partial_\n \p_2  \nonumber\\
		&& - \frac18 \bar\p_1 \g_\n \partial_\m \p_2 -  \frac18 \bar\p_2 \g_\m \partial_\n \p_1 +  \frac18 \bar\p_2  \g_\n \partial_\m \p_1  \,. \qquad 
	\end{eqnarray}
	These composite expressions indicate that the lowest-order Lagrangian, which contains only time-derivatives given as $\mathcal{O}(c^{-4})$, identically vanishes as $F_{00} = 0$. Once the fields are rescaled in accordance with \eqref{Scaling}, the lowest-order composite expression for $F_{\mu\nu}$ and $\vf$ are given by
	\begin{align}
		\varphi & = \frac{c^{1/2}}{8} (  \g^0 \dot\f_1 \p_2  -  \g^0 \dot\f_2 \p_1  - \frac12 f_1 \p_2 + \frac12 f_2 \p_1 ) \,,\nonumber\\
		F_{0i} & = \frac{1}{8}(\bar\p_2 \gamma_i \dot\p_1 - \bar\p_1 \gamma_i \dot\p_2)  \,, \quad  F_{ij}  = \mathcal{O}(c) \,.
	\end{align}
	Most importantly, the lowest order $F_{0i}$ does not contain any purely bosonic part as they are included in $\mathcal{O}(c)$ due to rescaling of the fields. Consequently, the Carroll limit gives rise to the following supersymmetric model consisting of four-fields interactions
	%%%%%%%%%%%%%%%%%
	\be
	\cL_{4f} = \frac12 \alpha b_i b^i + 2 \alpha \bar\l  \g_0 \dot\l\,,
	\label{TheModel}
	\ee
	where
	\begin{align}
		b_i &= \frac{1}{8} (\bar\p_2 \gamma_i \dot\p_1 - \bar\p_1 \gamma_i \dot\p_2) \,, \nonumber\\
		\l &= \frac{1}{8} (  \g^0 \dot\f_1 \p_2  -  \g^0 \dot\f_2 \p_1  - \frac12 f_1 \p_2 + \frac12 f_2 \p_1 ) \,.
	\end{align}
	%%%%%%%%%%%%%%%%
	This is an example of a supersymmetric model that contains only purely fermionic part and mixing terms, but no purely bosonic part. The supersymmetry of the model can also be checked explicitly by noticing the transformation rules for $b^i$ and $\lambda$ are given by $\delta b_i  = \bar\epsilon \gamma_i \dot{\lambda}$ and $\delta\l = \frac14 \, b^i \gamma_{i0} \e$. Upon using the transformation rules, the Lagrangian \eqref{TheModel} is invariant up to a total derivative term $\partial_t ( \bar\epsilon \gamma_i \lambda b^i)/2$. 
	
	%\textit{Higher-Order Corrections to $\mathcal{N} = 2$ SYK Model} --    In this section, we consider a 
	As mentioned, the Carrollian supersymmetric model without a bosonic part \eqref{TheModel} does not contain a spatial derivative, which enables us to calculate a higher-order correction to $\mathcal{N} =2$ SYK model \cite{Anninos:2016szt,Biggs:2023mfn} by reducing \eqref{TheModel} to $(0+1)$ dimensions. This is achieved by first generalizing our construction to $N$-number of scalar multiplets with a cubic potential term. In this case, the two-derivative action is given by
	\begin{eqnarray}
		\cL_{NS} &=& - \frac12 \partial_\m \f^A \partial^\m \f^A - \frac18 \bar\p^A \slashed\partial \p^A + \frac18 f^A f^A \nonumber\\
		&& + C_{ABC} \big( \f^A \f^B f^C + \frac12 \f^A \bar\p^B \p^C \big)  \,,
	\end{eqnarray}
	where $C_{ABC}$ is fully symmetric in its indices and $A,B = 1,2,\ldots,N$ counts the number of multiplets. The higher-derivative part is still given by the vector multiplet action \eqref{VectorAction}, but with the following composite expressions for $(F_{\m\n},\,\varphi)$
	\begin{eqnarray}
		\vf &=& \frac18 C_{AB} \big(\slashed\partial \f^A \p^B - \frac12 f^A \p^B \big) \,,\nonumber\\
		F_{\m\n} &=& C_{AB} \big(\partial_\m \f^A \partial_\n \f^B + \frac14 \bar\p^A \gamma_{[\m} \partial_{\n]} \p^B \big) \,,
	\end{eqnarray}
	where $C_{AB}$ is antisymmetric in its indices. Upon rescaling the fields in accordance with \eqref{Scaling} and $C_{ABC}$ with a factor of $c^{-2}$, and finally taking the $c \to 0$ limit, we obtain the Carrollian supersymmetric $N$ scalar multiplet model 
	\begin{eqnarray}
		\cL_{CS}& =&\frac12 \dot\f^A \dot\f^A + \frac18 \bar\p^A \g_0 \dot\p^A + \frac18 f^A f^A + C_{ABC} \f^A\f^B f^C \nonumber\\
		&& +\frac12 C_{ABC} \f^A\bar\p^B \p^C + \frac12 \alpha b_i b^i + 2 \alpha  \bar\l  \g_0 \dot\lambda \,.
		\label{extsyk}
	\end{eqnarray}
	Here, the composite expressions for $b_i$ and $\l$ are given by
	\begin{eqnarray}
		b_i &=&\frac18 C_{AB} \bar\p^A \g_i \dot \p^B \,,\nonumber\\
		\l_i &=& \frac18 C_{AB} \dot\f^A \g_0 \p^B_i - \frac12 C_{AB} f^A  \p^B_i \,.
	\end{eqnarray}
	This model is invariant under the following set of transformation rules 
	\bea
	\d f^A & =& \frac12 \bar\e \g_0 \dot\p^A \,, \qquad  \d \f^A = \frac14 \bar\e \p^A \,, 
	\nonumber\\
	\d \p^A & =& - \g_0 \dot \f^A \e - \frac12 f^A \e \,.
	\label{trans3}
	\eea
	Here, since the Lagrangian \eqref{extsyk} does not contain any spatial derivative, it can be viewed as a collection of infinitely many identical (0+1)-dimensional model labeled by $(x, y)$. 
	Choosing gamma-matrices as $\gamma_0 = \rmi \s_2, \g_1 = \s_1, \g_2 = \s_3$, the $(0+1)$-dimensional model is given by
	\begin{eqnarray}
		\cL_{\rm ESYK} &=& \frac12 \dot\f^A \dot\f^A  + \frac18 f^A f^A + \frac{\rmi}{8}  \big( \p_1^A \dot \p_1^A + \p_2^A \dot \p_2^A \big) \nonumber\\
		&& + C_{ABC} \left(\f^A \f^B f^C + \rmi \f^A \p_1^B \p_2^C \right) \nonumber\\
		&& + \frac12 \a \big(b_1^2 + b_2^2 + 4\rmi \l_1 \dot \l_1 + 4 \rmi  \l_2 \dot \l_2\big) \,,
		\label{1dModel}
	\end{eqnarray}
	which is a new higher derivative extension of ${\cal N}=2$ SYK-like model. The subscript $1,2$ refer to the components of the 3D spinor, i.e., $\psi = (\psi_1, \psi_2)$ and we define
	\begin{eqnarray}
		b_1 &=& \frac{\rmi} 8  C_{AB} \big(\p_1^A \dot \p_1^B - \p_2^A \dot \p_2^B \big) \,, \nonumber\\
		b_2 & = & - \frac{\rmi}{8}  C_{AB} \big( \p_1^A \dot \p_2^B +  \p_2^A \dot \p_1^B \big) \,,\nonumber\\
		\lambda_1 &=& \frac18 C_{AB} \dot \f^A \p_2^B - \frac12 C_{AB} f^A \p_1^B \,,\nonumber\\
		\lambda_2 &=& - \frac18 C_{AB} \dot \f^A \p_1^B - \frac12 C_{AB} f^A \p_2^B \,.
	\end{eqnarray}
	This model is invariant under the following set of off-shell transformation rules obtained by recasting the 3D transformation rules \eqref{trans3} in terms of (0+1)-dimensional variables
	\begin{align}
		\d \f^A &= \frac{\rmi}{4} \big( \e_1 \p_2^A - \e_2 \p_1^A \big) \,, & \d \p_1^A &= - \dot \f^A \e_2 - \frac12 f^A \e_1\,, \nonumber\\
		\d f^A &= \frac{\rmi}{2} \big(\e_1 \dot \p_1^A + \e_2 \dot \p_2^A \big) \,, & \d \p_2^A &= \dot \f^A \e_1 - \frac12 f^A \e_2 \,, 
		\label{CompositeSYK}
	\end{align}
	The higher-derivative part of the Lagrangian \eqref{1dModel}, which does not contain any purely bosonic terms, represents an example of an off-shell extension of the $\mathcal{N}=2$ SYK-like model. It is worthwhile to mention that the higher derivative Lagrangian \eqref{TheModel}
	resembles the $\mathcal{N}=1$ SYK model for two vector multiplets with $C_{ABC}=0$ studied in \cite{Fu:2016vas}, if the composite objects $(b_i, \l_i)$ are viewed as fundamental fields. Thus, our higher-derivative SYK-like action is in fact the sum of SYK (built out of composite fields) and SYK-like models.
	
	In this work, we show that the Carrollian supersymmetry has the curious feature that a supersymmetric Lagrangian does not require a purely bosonic part. By taking Carrollian limit of relativistic models we are able to provide a concrete supersymmetric example with such an exotic feature. A better understanding might be possible by constructing a Carrollian superspace in which case various supersymmetric results could easily be obtained and their mathematical structure can be better studied. As the 3D example given here does not contain spatial derivatives, it is straightforward to relate it to a SYK-like model with ${\cal N}=2$ off-shell supersymmetry extended by higher-order interactions which lacks a purely bosonic contribution. Recently a curious low energy behavior of ${\cal N}=2$ SYK-like model,  was observed in \cite{Biggs:2023mfn} namely the entropy $S\approx S_0 + (\text{const.}) T^a$ with $a \neq 1$. By dimensional analysis, the higher order terms constructed here are expected to modify this result by a term $\alpha T$ which may indicate the existence of new phases. The higher-order terms however will not generate propagating ghosts in the trivial vacuum with $\phi^A=f^A=0$ since they are at least quartic in fields. Whether they imply instability at
	fully nonlinear level requires a proper quantization of the Carroll invariant models that is still under development.
	
	Another intriguing aspect is that, as we have shown, 
	%the ultra-relativistic sector of a supersymmetric vector multiplet action is in fact a collection of %infinitely many SYK models, each labelled by its spatial coordinates. In our 3D example, 
	the SYK-like model descending from a composite 3D ${\cal N}=1$ supersymmetric vector multiplet action lacks the cubic interaction term characterized by $C_{ABC}$ as
	$\bar\lambda \gamma \cdot F \lambda$ vanishes identically in $D=3$ due to the symmetry of gamma matrices. However, such a term does exist in $D \geq 4$ \cite{Freedman:2012zz}, which indicates that along our procedure, their ultra-relativistic sector should lead to a complete SYK model based on the composite fields, with the cubic interaction term. It should also be interesting to investigate effects of the higher order terms on various physical quantities in  ${\cal N}=2$ model extending earlier results on this subject \cite{Fu:2016vas, Peng:2017spg, Murugan:2017eto, Biggs:2023mfn}.  Different from the top-down approach by taking the $c\to 0$ limit of relativistic models, one can adopt another approach by considering the spinor representations of the homogeneous Carroll group based on the degenerate Clifford algebra. This bottom-up approach enables the construction of intrinsic supersymmetric Carrollian models without an obvious relativistic origin.

	Our investigation suggests a potential connection between Carrollian supersymmetric field theory and Jackiw-Teitelboim (JT) supergravity, resonating with the duality between JT gravity and the SYK model. Furthermore, considering the significance of Carrollian field theory in celestial holography \cite{Donnay:2022aba}, it hints at a deeper relationship between JT supergravity (possibly infinite copies) and celestial holography.

	%\textit{Conclusions and outlook.}---
	
	\textit{Acknowledgements.}---   We are grateful to E. Bergshoeff, J. Distler,  J. Rong and N. Su for useful communications. M.O. and U.Z. are supported in part by TUBITAK grant 121F064. M.O. acknowledges the support by the Outstanding Young Scientist Award of the Turkish Academy of Sciences (TUBA-GEBIP). The work of Y.P. is supported by the National Key Research and Development Program under grant No. 2022YFE0134300 and by the National Natural Science Foundation of China (NSFC) under grant No. 12175164.

	\bibliographystyle{utphys}
	\bibliography{ref}

\end{document}